\documentclass[number,sort&compress]{elsarticle}

\usepackage{graphicx}
\usepackage{amsmath}   
\usepackage{lineno}
\nolinenumbers

\begin{document}

\begin{frontmatter}

\title{TORCH: a large area time-of-flight detector for particle identification}

\author[add0]{N.~Harnew\corref{cor}}
\ead{Neville.Harnew@physics.ox.ac.uk}
\author[add1,add2]{S.~Bhasin}
\author[add5]{T.~Blake}
\author[add1]{N.H.~Brook}
\author[add6]{T.~Conneely}
\author[add2]{D.~Cussans}
\author[add3]{M.~van~Dijk}
\author[add3]{R.~Forty}
\author[add3]{C.~Frei}
\author[add4]{E.P.M.~Gabriel}
\author[add0]{R.~Gao}
\author[add5]{T.J.~Gershon}
\author[add3]{T.~Gys}
\author[add0]{T.~Hadavizadeh}
\author[add0]{T.H.~Hancock}
\author[add5]{M.~Kreps}
\author[add6]{J.~Milnes}
\author[add3]{D.~Piedigrossi}
\author[add2]{J.~Rademacker}

\cortext[cor]{Corresponding author}

\address[add0]{Denys Wilkinson Laboratory, University of Oxford, Keble Road, Oxford OX1 3RH, United Kingdom}
\address[add1]{University of Bath, Claverton Down, Bath BA2 7AY, United Kingdom.}
\address[add2]{H.H. Wills Physics Laboratory, University of Bristol, Tyndall Avenue, Bristol BS8 1TL, United Kingdom}
\address[add3]{European Organisation for Nuclear Research (CERN), CH-1211 Geneva 23, Switzerland}
\address[add4]{School of Physics and Astronomy, University of Edinburgh, James Clerk Maxwell Building, Edinburgh EH9 3FD, United Kingdom }
\address[add5]{Department of Physics, University of Warwick, Coventry, CV4 7AL, United Kingdom}
\address[add6]{Photek Ltd., 26 Castleham Road, St Leonards on Sea, East Sussex, TN38 9NS, United Kingdom}

\begin{abstract}
TORCH  is a time-of-flight detector that is being developed for the Upgrade II of the LHCb experiment, with the aim of providing charged particle 
identification over the momentum range 2--10\,GeV/$c$. 
A small-scale TORCH demonstrator with customised readout electronics 
has been operated successfully in beam tests at the CERN PS. 
Preliminary results indicate that a single-photon resolution better than 100\,ps can be achieved.

\end{abstract}


\end{frontmatter}

\section{Introduction}

\begin{figure}
\centering
\includegraphics[width=0.41\linewidth]{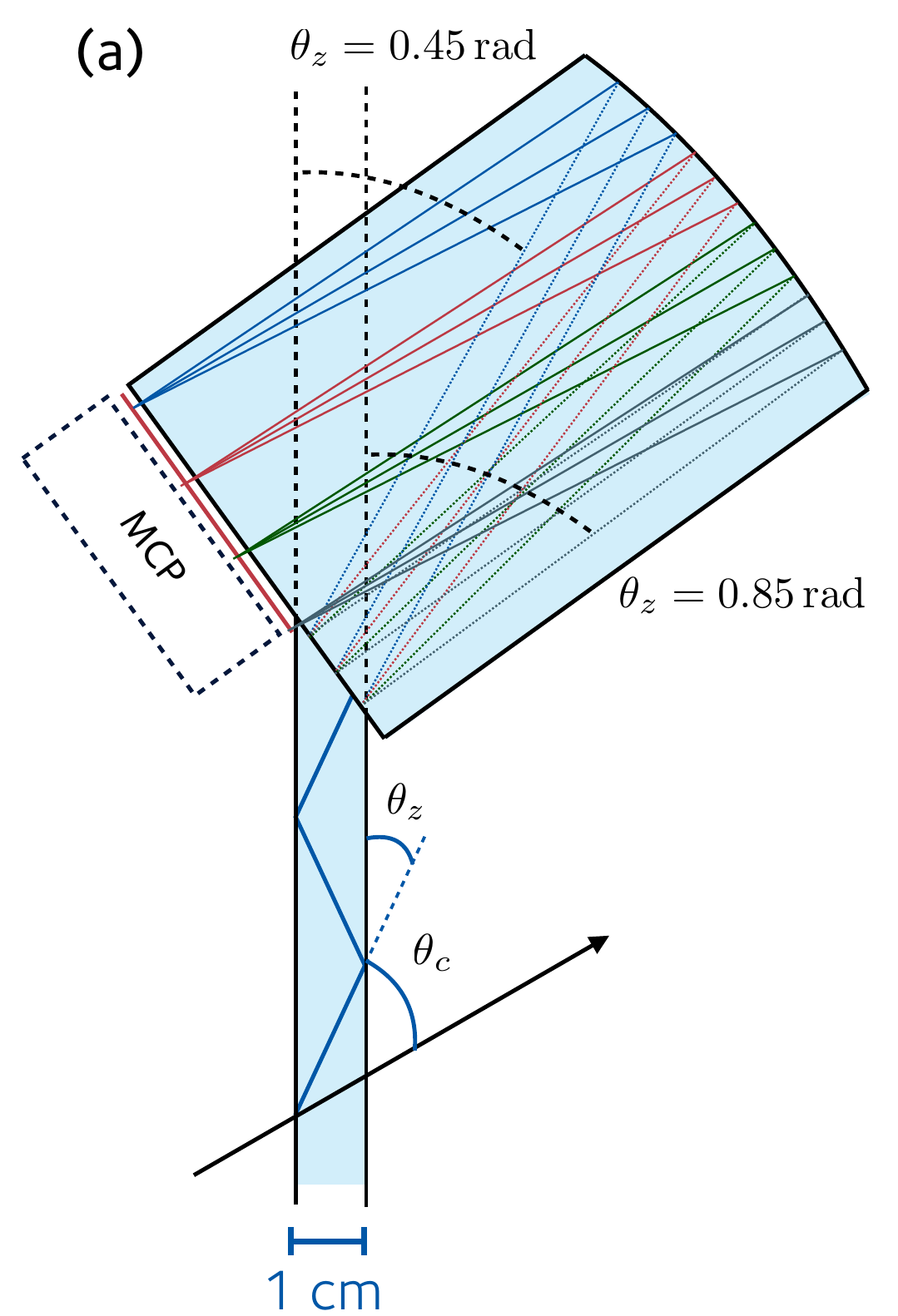}
\hspace{0.7truecm}
\includegraphics[width=0.35\linewidth]{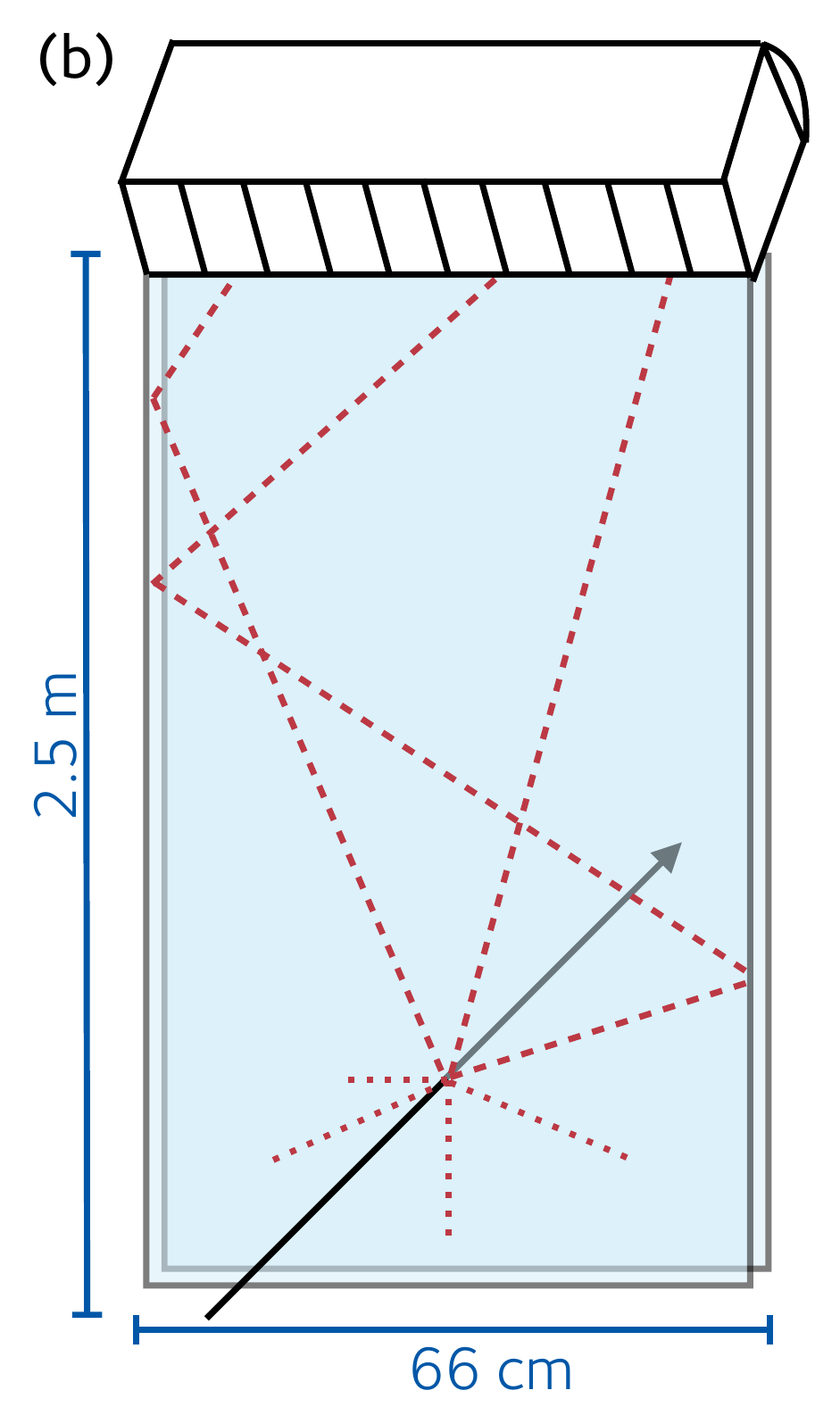}
\caption{Schematics  of a TORCH module showing possible reflection paths: (a) the focussing block and MCP plane, (b) a single  LHCb module.}
\label{fig:TORCH-schematics}
\end{figure}

TORCH is a time-of-flight detector designed to provide Particle IDentification (PID)
for low momentum particles between 2-10 GeV/$c$\,\cite{Charles:TORCH, Brook:TORCH}.
 The detector exploits prompt Cherenkov light   produced 
when charged particles traverse a  10\,mm thick quartz plate. 
The photons propagate via total internal reflection to the edge of the plate
where they are focussed onto a detector plane comprising 
micro-channel plate photomultipliers (MCP-PMTs), shown in Fig.\,\ref{fig:TORCH-schematics}(a). 
The aim is to measure individual photons  
with an angular precision of 1\,mrad and  70\,ps time resolution and, with $\sim$30 detected
photons,  this would result in $\sim$13\,ps per incident charged particle.

TORCH is proposed for the upgrade II of the LHCb experiment, where 
PID of pions, kaons and protons  is essential for CP violation measurements, exotic spectroscopy and particle
tagging\,\cite{LHCb:2017-EOI-upgrade}.   
To provide the necessary PID discrimination up to 10 GeV/$c$ 
over the full spectrometer acceptance, TORCH forms a time of flight wall of area 5 $\times$ 6\,m$^2$, 
located 10\,m from the proton-proton interaction region.
The detector is divided into 18 modules, each 66\,cm wide and 2.5\,m high, illustrated in Fig.\,\ref{fig:TORCH-schematics}(b). 
To meet the challenging LHC environment, TORCH is designed for high occupancy 
and significant  pile-up.

When a charged particle passes through a TORCH module, 
the track entry point in the quartz radiator is measured from 
position information provided by  the  tracking system of the LHCb
spectrometer.  The characteristic photon patterns are reconstructed 
from the MCP-PMT photon detectors and  corrections  made to account for chromatic dispersion.
To then distinguish the different particle species,
the expected distributions for $\pi /K /p$  hypotheses are 
compared to the measured MCP-PMT photon spatial hits and arrival times.

\section{Micro-channel plate PMTs}

To achieve the required angular resolution of TORCH,  each MCP-PMT detector
requires 128 $\times$ 8 effective granularity in the transverse and longitudinal projections
over a 53 $\times$ 53\,mm$^2$ active area, with 11 MCP-PMTs per module.
The MCP-PMTs\,\cite{Conneely_2015_PSD_proceedings} are developed with industrial partners, Photek UK. The 
pixels are 64 $\times$ 64 and grouped to read out with 
64 $\times$ 8 granularity; 
charge sharing is then used to give required 128 granularity in the transverse direction.
 To survive the LHC environment, each MCP-PMT must also be 
capable of withstanding a large integrated charge on its anode 
(5 C\,cm$^{-2}$),
achieved using an atomic-layer deposition (ALD) coating\,\cite{ref:Conneely, Gys_2016_RICH_proceedings}.
The MCP-PMTs are read out with a customised electronics system\,\cite{Gao-2016-TWEPP2015-proceedings}, 
and utilize the NINO32\,\cite{Despeisse-2011-NINO32} and HPTDC\,\cite{Christiansen:HPTDC} chipsets, 
developed for fast timing applications of the ALICE experiment.

\begin{figure}
\centering
\hspace{0.35truecm}
\includegraphics[width=0.75\linewidth]{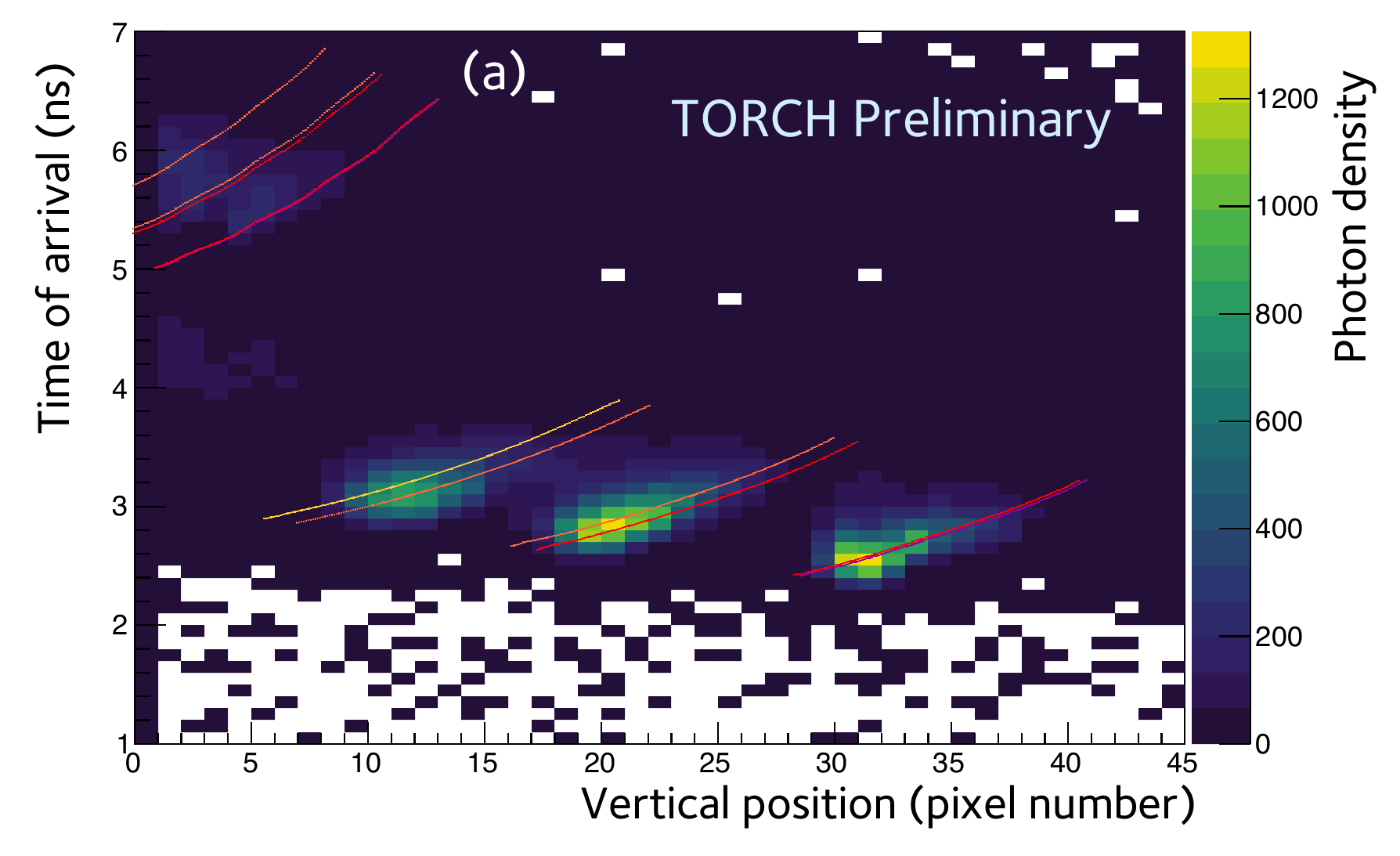}
\hspace{-0.5truecm}
\includegraphics[width=0.65\linewidth]{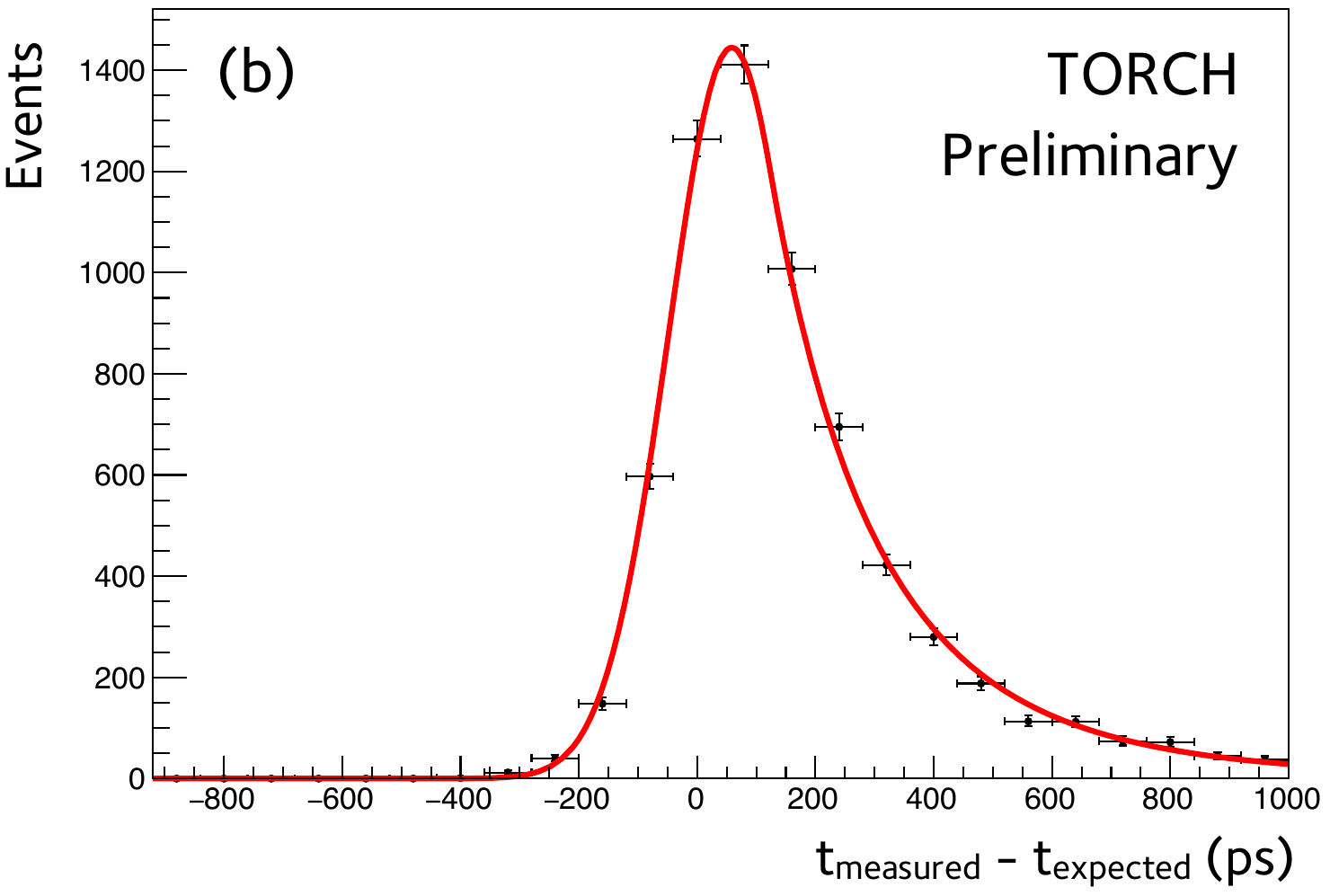}
\caption{(a) The time-of-arrival of single Cherenkov photons from a 5\,GeV/$c$ pion beam,  
relative to a beam time-reference station,
as a function of detected vertical column pixel number. The overlaid lines
represent the simulated patterns for light reflected  only off the front and back  faces
of the radiator plate  (lower right distribution),  light also undergoing one and two
reflections off the side faces (lower central and left  distributions, respectively), 
and multiple reflections from the bottom horizontal  face (top left distributions).
(b) The distribution of residuals between the measured Cherenkov photon arrival times 
with respect to the simulated distribution, with a ``crystal ball'' fit superimposed.}
\label{fig:timeprojection}
\end{figure}

\section{Beam tests}

A small-scale TORCH demonstrator module was tested in CERN PS 5\,GeV/$c$ mixed pion-proton 
beam in November 2017. 
The quartz radiator has dimensions 120 $\times$ 350 $\times$ 10\,mm$^3$
and  read out with a single Photek MCP-PMT of 64 $\times$ 4 granularity. 
The radiator plate was mounted in an almost vertical position, tilted backwards by 5$^\circ$ with respect to the horizontal  incidence of the beam. 
Pions and protons were distinguished by an external time-of-flight telescope.
A method of data-driven calibration was employed
 to correct simultaneously for time-walk
and integral non-linearities of the electronics\,\cite{Brook:TORCH}.

Data analysis from the test-beam run is  well advanced.
Characteristic photon  bands are observed due to reflections from the faces of the radiator plate.
With the beam impinging approximately 14\,cm below the plate centre and close to its side, 
the patterns show the expected distribution of reflections, shown in Fig.\,\ref{fig:timeprojection}\,(a). 
The residuals between the observed and simulated times of arrival  are shown in  Fig.\,\ref{fig:timeprojection}\,(b),
resulting in a resolution per photon of $\sim$100\,ps.
The tail in the distribution is caused by imperfect corrections, and improvements are expected
when charge-to-width calibrations are incorportated. 

\section{Summary and future plans}

A  single-photon timing resolution of approximately 100\,ps has been achieved with a small-scale TORCH demonstrator 
utilising  a customised 64 $\times$ 4 pixelated MCP-PMT, and the timing performance
is approaching that required for the Upgrade II of the LHCb experiment. 
This prototype is a precursor to a half-length
TORCH module (660 $\times$ 1250 $\times$ 10\,mm$^3$)  which is currently under construction. 
The half-length module will be equipped with ten 64 $\times$ 8 pixel MCP-PMTs and 
will be ready for test-beam operation at the end of 2018.

\section*{Acknowledgments}

The support is acknowledged of the STFC Research Council, UK, and of the 
European Research Council  through an FP7
Advanced Grant  (ERC-2011-AdG 299175-TORCH).



\begin{thebibliography}{10}
\expandafter\ifx\csname url\endcsname\relax
  \def\url#1{\texttt{#1}}\fi
\expandafter\ifx\csname urlprefix\endcsname\relax\def\urlprefix{URL }\fi



\bibitem{Charles:TORCH}
M.~Charles and R.~Forty: TORCH: Time of flight identification with
  Cherenkov radiation, Nucl. Instrum. Methods A639 (2011) 173 -- 176.

\bibitem{Brook:TORCH}
N.~Brook et~al., Testbeam studies of a TORCH prototype detector,
submitted to Nucl. Instrum. Methods, 
  \urlprefix\url{https://arxiv.org/abs/1805.04849}.

\bibitem{LHCb:2017-EOI-upgrade}
The LHCb Collaboration, Expression of Interest for a Phase-II LHCb Upgrade: 
Opportunities in flavour physics, and beyond, in the HL-LHC era, CERN-LHCC-2018-003 (2017).


\bibitem{Conneely_2015_PSD_proceedings}
T.~Conneely et~al., The TORCH PMT: a close packing, multi-anode, long
  life MCP-PMT for Cherenkov applications,  Journal of Instrumentation
  vol.~10 (2015), no.~05: C05003.

\bibitem{ref:Conneely} T. M. Conneely, J. S. Milnes, J. Howorth (Photek Ltd), Characterisation and lifetime 
measurements of ALD coated microchannel plates in a sealed photomultiplier tube, 
Nucl. Instr. and Meth.,  A732 (2013) 388--391.

\bibitem{Gys_2016_RICH_proceedings}
T.~Gys et al.,  The TORCH detector R\&D: Status and perspectives, 
Nucl. Instrum. Methods  A876 (2017)  156 -- 159.

\bibitem{Gao-2016-TWEPP2015-proceedings}
R.~Gao et~al., Development of TORCH readout electronics for customised MCPs,
Journal of Instrumentation  11 (2016)  C04012.


\bibitem{Despeisse-2011-NINO32}
M.~Despeisse, F.~Powolny, P.~Jarron and J.~Lapington,  Multi-Channel
  Amplifier-Discriminator for Highly Time-Resolved Detection, Nuclear
  Science, IEEE Transactions  58 (2011) 202 -- 208.


\bibitem{Christiansen:HPTDC} J. Christiansen,  High Performance Time to Digital Converter, CERN/EP-MIC (2002).




\end{thebibliography}
\end{document}